\documentclass[submitted]{ieeeaccess}
\usepackage{amsmath,amssymb,amsfonts}
\usepackage{changepage}
\usepackage{cite}
\usepackage{graphicx}
\usepackage{textcomp}
\usepackage{adjustbox}

\usepackage{color,soul}
\newcommand{\DeletedText}[1]{}
\newcommand{\NewCorrection}[1]{{#1}}

\begin{document}
    
%\history{Date of publication xxxx 00, 0000, date of current version xxxx 00, 0000.}
%\doi{10.1109/ACCESS.2020.DOI}

\history{Submitted}
\doi{}

\title{LUDB: a new open-access validation tool for electrocardiogram delineation algorithms}

\author{%
    \uppercase{Alena I. Kalyakulina}\authorrefmark{1},
    \uppercase{Igor I. Yusipov}\authorrefmark{1},
    \uppercase{Victor A. Moskalenko}\authorrefmark{1},
    \uppercase{Alexander V. Nikolskiy}\authorrefmark{2},
    \uppercase{Konstantin A. Kosonogov}\authorrefmark{2},
    \uppercase{Grigory V. Osipov}\authorrefmark{1},
    \uppercase{Nikolai Yu. Zolotykh}\authorrefmark{1*},
    \uppercase{Mikhail V. Ivanchenko}\authorrefmark{1}%
}

\address[1]{Institute of Information Technologies, Mathematics and Mechanics, Lobachevsky University, Nizhni
Novgorod, Russia}

\address[2]{Department of Cardiovascular Surgery, City Clinical Hospital No 5, Nizhni Novgorod, Russia}

\corresp{Corresponding author: Nikolai Yu. Zolotykh (e-mail: nikolai.zolotykh@itmm.unn.ru, orcid: 0000-0003-4542-9233).}

\begin{abstract}
We report Lobachevsky University Database (LUDB) of ECG signals, \NewCorrection{an open tool for validating ECG delineation algorithms, that is superior to the existing publicly available data bases in several aspects. LUDB} contains $200$ recordings of $10$-second $12$-lead electrocardiograms (ECG) from different subjects, \NewCorrection{representative of a variety of signal morfologies}. The boundaries and peaks of QRS \NewCorrection{complexes} and P and T waves are manually annotated by cardiologists for all recordings and independently for each lead, and all records received an expert classification by abnormalities. \DeletedText{In addition, the database is representative of a variety of signal morphologies. These features make LUDB a promising tool for validating ECG delineation algorithms across a broad range of ECG signal shapes and patient diagnoses.} \NewCorrection{We present} a case study for the recently proposed wavelet-based algorithm \DeletedText{is presented.} \NewCorrection{and the broadly used ecg-kit tool, and demonstrate the advantage of multi-lead ECG data analysis. LUDB contributes to the diversity of public databases employed in developing and validating novel ECG analysis algorithms, including the most advanced based on deep learning neural networks.}  
\end{abstract}


\tfootnote{This work was supported by the Ministry of Science and Higher Education of the Russian Federation, Agreements No. 074-02-2018-330\,(1) and No. 13.1902.21.0026.}

\titlepgskip=-15pt

\maketitle

\section*{Introduction}\label{introduction}

\begin{figure}
    \centering
\includegraphics[width=85mm]{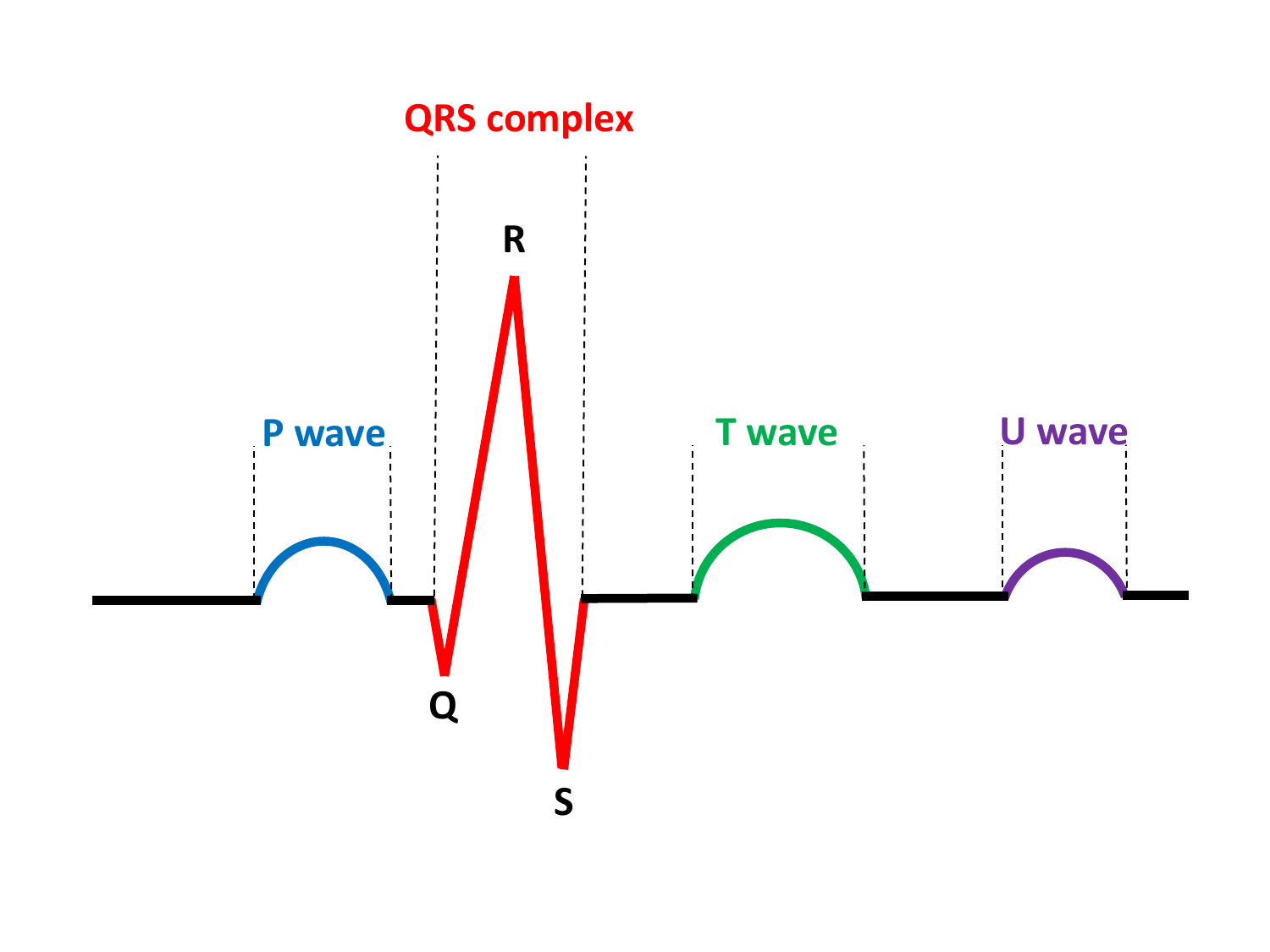}
\caption{ECG signal. Schematic representation of \NewCorrection{the QRS complex and P, T, U waves. U wave may not always be observed due to its small size. Its exact source remains unclear.}}
\label{fig1}
\end{figure}

Recording the electrical activity of heart, or electrocardiography, is one of the basic medical diagnostic means for assessing cardiac activity, in particular, determining the heart rate and rhythm disturbances.
The voltage graphs -- electrocardiograms (ECGs) manifest repeated activity with the commonly identified structural elements of each heart beat image: QRS complex, P and T waves (Fig. \ref{fig1}). Analysis of their amplitudes, shapes (morphologies) and durations allows for identifying cardiac rhythm disorders and cardiovascular diseases, such as ischemia and myocardial infarction \cite{Khan2009}. A rich variety of signal morphology, accompanied by their non-stationary nature, potential defects in recordings and noise, makes an automated search for these waves and complexes, also known as ECG delineation \NewCorrection{(also known as ECG segmentation or ECG annotation)}, a challenging task.

This problem has been tackled for quite a while, resulting in a number of algorithms that solve it at different level of detail. The first ones were designed to detect the QRS complex only, referring on the amplitude of the ECG signal and its first derivative \cite{Pan1985}. Detecting boundaries and peaks of P and T waves required more sophisticated methods based on wavelet transform \cite{Li1995, Martinez2004}, Hilbert transform \cite{Benitez2001}, phasor transform \cite{Martinez2010}, hidden Markov models \cite{Graja2005}, gradient based algorithms \cite{Mazomenos2012} and morphological transforms \cite{Sun2005}.

Validating delineation algorithms requires standardized datasets with complexes and waves that are manually annotated by specialists. Increasing their number and variety is crucial itself, for both better training and testing robustness of developed methods. Moreover, several collections that are currently available in the public domain: MIT-BIH Arrhythmia Database \cite{mitdb}, European ST-T Database \cite{edb}, and QT Database \cite{qtdb}, have certain limitations. That is, MIT-BIH Arrhythmia Database and European ST-T Database have a markup only for QRS complexes. In turn, the QT Database contains annotations for P, QRS and T waves, but has only 2-lead Holter recordings, and is, therefore, not suitable for validating multilead delineators, which are currently the most common approach.

ECG database assembled at Lobachevsky University (LUDB) is free from these issues. The reported database consists of $200$ recordings of standard $10$-second $12$-lead recordings \cite{IEC} from different subjects, representing a variety of signal morphologies. The boundaries of P, QRS and T complexes at each lead are manually annotated by cardiologists for all $200$ records, and each subject is supplemented with noticed abnormalities (same as in the other studies, we skip U-wave due to its small amplitude and noise issues). The overall number of annotated complexes in LUDB considerably exceeds that in QTDB. Altogether, these features make LUDB a valuable contribution to the current publicly available sources. 

As the case study, we made use of this dataset for validating our recent algorithm \cite{Kalyakulina2018}, that implements wavelet transform for multi-lead multi-morphology analysis with error correction, and make a comparison to the popular ecg-kit tool \cite{ECGKit}, which employs one of its predecessors, a single-lead delineator \cite{Martinez2004}. Expectedly, the results demonstrate a comparable performance of both for \NewCorrection{QTDB} and a noticeable improvement of delinearing P and T waves for LUDB \NewCorrection{achieved} by the former algorithm. 

\NewCorrection{We note that there are many recent studies related to the ECG processing including disease detection, delineation, sleep staging, biometric human identification, denoising, and others (see recent overview \cite{Hong2020}). In this article, we only focus on the task of ECG delineation. 
The solution to this task can be used to solve other problems, in particular, the disease detection. On the other hand, using standard annotations and expert features not always be the best choice. Automatically generated features (such as deep learning features) can be more informative than the expert features. In particular, there have been noticeable successes in the problem of automatic recognition of cardiac diseases using sparse representation of ECG \cite{RajRay2018}, using deep learning generated features \cite{Hong2017,Hong2019}, combination of artificial intelligence methods and linear and non-linear decomposition \cite{Abdalla2019}, different feature extraction methods with machine learning algorithms \cite{Marinho2019}, different end-to-end ECG deep learning classifiers, e.g. \cite{Xu2018,Hannun2019}, etc.}

The paper is organized as follows.
In Section~\ref{database}, we describe the LUDB database.
Section~\ref{algorithm} contains an outline of the delineation algorithm \cite{Kalyakulina2018}. A case study of its validation with LUDB and QTDB is reported in Section~\ref{results}.
Section~\ref{conclusions} summarizes the results and perspectives.

\section{Lobachevsky University Database}\label{database}

A publicly available Lobachevsky University Database \cite{Our_PhysioNet} contains $200$ records from $200$ subjects in wfdf format \cite{PhysioNet}. 

The ECGs were collected from healthy participants and patients of the Nizhny Novgorod City Hospital No.5 in the period 2017--2018 with various cardiovascular diseases, some of them had pacemakers. The records were made by specialized medical staff (functional diagnostics nurses).
All participants provided informed written consent before participating in the experiment. The age of subjects varied from 11 to 90 years, with the average 52 years, the distribution by gender was 85 women and 115 men. Table \ref{table2} reports the breakdown by the type of rhythm and Table \ref{table3} by the type of heart electrical axis. These parameters are specified for all records in the database.

ECG recordings were obtained by the Schiller Cardiovit AT-101 cardiograph \cite{schiller}, with conventional $12$ leads (I, II, III, aVR, aVL, aVF, V1, V2, V3, V4, V5, V6), the duration is $10$ seconds, the signals are digitized at $500$ Hz, complying with the international \NewCorrection{standard} \cite{IEC}. 

The boundaries and peaks of QRS \NewCorrection{complexes}, and P and T waves were determined by two certified and practicing cardiologists (A.V.\,N. and K.A.K.) by an eye inspection of each ECG signal and independently for each of 12 leads. 
The markup of all ECG forms was joint, relying on standard criteria \cite{book} and based on consensus opinion, as well as classification of abnormalities. This approach was chosen as to decrease subjective influence and provide the end user with a definite annotation. The recordings and markup files in the database come separately, and are open for download and further independent exploration, in particular, with regard to assessing variability in expert opinion. In total, the dataset contains 58429 annotated waves, that is almost six times greater than in the widely referred QT database (Table \ref{table1}), which is the only publicly available database with all the waves annotated, to the best of our knowledge.

Tables \ref{table4}, \ref{table5} summarize the content of the database by main ECG abnormalities and their count. Note that some patients would have several issues at the same time.

\begin{table}
	%\begin{adjustwidth}{-2.25in}{0in} % Comment out/remove adjustwidth environment if table fits in text column.
		\centering
		\caption{Comparative numbers of annotated waves in QTDB and LUDB}
		\medskip
		
		\begin{tabular}{|c|c|c|c|c|}
			\hline
			&&&&\\[-1em]
			& {P wave} & {QRS complex} & {T wave} & {Total} \\
			\hline
			&&&&\\[-1em]
			QTDB & 3194 & 3623 & 3542 & 10359 \\
			\hline
			&&&&\\[-1em]
			LUDB & 16797 & 21966 & 19666 & 58429 \\
			\hline
		\end{tabular}
		\label{table1}
	%\end{adjustwidth}
\end{table}

\begin{table}
	%\begin{adjustwidth}{-2.25in}{0in} % Comment out/remove adjustwidth environment if table fits in text column.
		\centering
		\caption{Breakdown in heart rhythm types, represented in the database}
		\medskip
		\begin{tabular}{|l|c|}
			\hline
			\multicolumn{1}{c|}{Rhythm} & {Number of subjects} \\
			\hline
			&\\[-1em]
			Sinus rhythm & 143 \\
			\hline
			&\\[-1em]
			Sinus tachycardia & 4 \\
			\hline
			&\\[-1em]
			Sinus bradycardia & 25 \\
			\hline
			&\\[-1em]
			Sinus arrhythmia & 8 \\
			\hline
			&\\[-1em]
			Irregular sinus rhythm & 2 \\
			\hline
			&\\[-1em]
			\NewCorrection{Atrial fibrillation}	 & \NewCorrection{15} \\
			\hline
			&\\[-1em]
			\NewCorrection{Atrial flutter, typical}	 & \NewCorrection{3} \\
			\hline
			&\\[-1em]
			{\it Total} & 200\\
			\hline
		\end{tabular}
		\label{table2}
	%\end{adjustwidth}
\end{table}

\begin{table}
	%\begin{adjustwidth}{-2.25in}{0in} % Comment out/remove adjustwidth environment if table fits in text column.
		\centering
		\caption{Breakdown in types of electrical axis, represented in the database}
		\medskip
		\begin{tabular}{|l|c|}
			\hline
			&\\[-1em]
			{Electric axis of the heart} &{Number of subjects} \\
			\hline
			&\\[-1em]
			Normal & 75 \\
			\hline
			&\\[-1em]
			Left axis deviation	 & 66 \\
			\hline
			&\\[-1em]
			Vertical & 26 \\
			\hline
			&\\[-1em]
			Horizontal & 20 \\
			\hline
			&\\[-1em]
			Right axis deviation & 3 \\
			\hline
			&\\[-1em]
			Undetermined & 10 \\
			\hline
			&\\[-1em]
			{\it Total} & 200\\
			\hline
		\end{tabular}
		\label{table3}
	%\end{adjustwidth}
\end{table}

\begin{table}
	%\begin{adjustwidth}{-2.25in}{0in} % Comment out/remove adjustwidth environment if table fits in text column.
		\centering
		\caption{Breakdown in cardiovascular disorders, represented in the database (conduction abnormalities, extrasystole, hypertrophy, cardiac pacing)}
		\medskip
        \begin{adjustbox}{width=1\linewidth}
    		\begin{tabular}{|l|c|}
			\hline
			&\\[-1em]
			\multicolumn{1}{c|}{Conduction abnormalities} & {Number of subjects} \\
			\hline
			&\\[-1em]
			Sinoatrial blockade, undetermined & 1 \\
			\hline
			&\\[-1em]
			I degree AV block & 10 \\
			\hline
			&\\[-1em]
			III degree AV-block & 5 \\
			\hline
			&\\[-1em]
			Incomplete right bundle branch block  & 29 \\
			\hline
			&\\[-1em]
			Incomplete left bundle branch block & 6 \\
			\hline
			&\\[-1em]
			Left anterior hemiblock & 16 \\
			\hline
			&\\[-1em]
			Complete right bundle branch block & 4\\
			\hline
			&\\[-1em]
			Complete left bundle branch block & 4 \\
			\hline
			&\\[-1em]
			Non-specific intravintricular conduction delay & 4 \\
			\hline
			&\\[-1em]
			\multicolumn{1}{c|}{Extrasystole} &{Number of subjects} \\
			\hline
			&\\[-1em]
			Atrial extrasystole: undetermined & 2 \\
			\hline
			&\\[-1em]
			Atrial extrasystole: low atrial & 1 \\
			\hline
			&\\[-1em]
			Atrial extrasystole: left atrial & 2 \\
			\hline
			&\\[-1em]
			Atrial extrasystole: SA-nodal extrasystole & 3 \\
			\hline
			&\\[-1em]
			Atrial extrasystole, type: single PAC & 4 \\
			\hline
			&\\[-1em]
			Atrial extrasystole, type: bigemini & 1 \\
			\hline
			&\\[-1em]
			Atrial extrasystole, type: quadrigemini & 1 \\
			\hline
			&\\[-1em]
			Atrial extrasystole, type: allorhythmic pattern & 1 \\
			\hline
			&\\[-1em]
			Ventricular extrasystole, morphology: polymorphic & 2 \\
			\hline
			&\\[-1em]
			Ventricular extrasystole, localisation: RVOT, anterior wall & 3 \\
			\hline
			&\\[-1em]
			Ventricular extrasystole, localisation: RVOT, antero-septal part & 1 \\
			\hline
			&\\[-1em]
			Ventricular extrasystole, localisation: IVS, middle part & 1 \\
			\hline
			&\\[-1em]
			Ventricular extrasystole, localisation: LVOT, LVS & 2 \\
			\hline
			&\\[-1em]
			Ventricular extrasystole, localisation: LV, undefined & 1 \\
			\hline
			&\\[-1em]
			Ventricular extrasystole, type: single PVC & 6 \\
			\hline
			&\\[-1em]
			Ventricular extrasystole, type: intercalary PVC & 2 \\
			\hline
			&\\[-1em]
			Ventricular extrasystole, type: couplet & 2 \\
			\hline
			&\\[-1em]
			\multicolumn{1}{c|}{Hypertrophy} &{Number of subjects} \\
			\hline
			&\\[-1em]
			Right atrial hypertrophy & 1 \\
			\hline
			&\\[-1em]
			Left atrial hypertrophy & 102 \\
			\hline
			&\\[-1em]
			Right atrial overload & 17 \\
			\hline
			&\\[-1em]
			Left atrial overload & 11 \\
			\hline
			&\\[-1em]
			Left ventricular hypertrophy & 108 \\
			\hline
			&\\[-1em]
			Right ventricular hypertrophy & 3 \\
			\hline
			&\\[-1em]
			Left ventricular overload & 11 \\
			\hline
			&\\[-1em]
			\multicolumn{1}{c|}{Cardiac pacing} &{Number of subjects} \\
			\hline
			&\\[-1em]
			UNIpolar atrial pacing & 1 \\
			\hline
			&\\[-1em]
			UNIpolar ventricular pacing & 6 \\
			\hline
			&\\[-1em]
			BIpolar ventricular pacing & 2 \\
			\hline
			&\\[-1em]
			Biventricular pacing & 1 \\
			\hline
			&\\[-1em]
			P-synchrony & 2 \\
			\hline
		\end{tabular}
		\label{table4}
        \end{adjustbox}
    	%\end{adjustwidth}
\end{table}

\begin{table}
	%\begin{adjustwidth}{-2.25in}{0in} % Comment out/remove adjustwidth environment if table fits in text column.
		\centering
		\caption{Breakdown in cardiovascular disorders, represented in the database (ischemia, repolarisation abnormalities)}
        \begin{adjustbox}{width=1\linewidth}
    		\begin{tabular}{|l|c|}
			\hline
			&\\[-1em]
			\multicolumn{1}{c|}{Ischemia} &{Number of subjects} \\
			\hline
			&\\[-1em]
			STEMI: anterior wall & 8 \\
			\hline
			&\\[-1em]
			STEMI: lateral wall & 7 \\
			\hline
			&\\[-1em]
			STEMI: septal & 8 \\
			\hline
			&\\[-1em]
			STEMI: inferior wall & 1 \\
			\hline
			&\\[-1em]
			STEMI: apical & 5 \\
			\hline
			&\\[-1em]
			Ischemia: anterior wall & 5 \\
			\hline
			&\\[-1em]
			Ischemia: lateral wall & 8 \\
			\hline
			&\\[-1em]
			Ischemia: septal & 4 \\
			\hline
			&\\[-1em]
			Ischemia: inferior wall & 10 \\
			\hline
			&\\[-1em]
			Ischemia: posterior wall & 2 \\
			\hline
			&\\[-1em]
			Ischemia: apical & 6 \\
			\hline
			&\\[-1em]
			Scar formation: lateral wall & 3 \\
			\hline
			&\\[-1em]
			Scar formation: septal & 9 \\
			\hline
			&\\[-1em]
			Scar formation: inferior wall & 3 \\
			\hline
			&\\[-1em]
			Scar formation: posterior wall & 6 \\
			\hline
			&\\[-1em]
			Scar formation: apical & 5 \\
			\hline
			&\\[-1em]
			Undefined ischemia/scar/supp.NSTEMI: anterior wall & 12 \\
			\hline
			&\\[-1em]
			Undefined ischemia/scar/supp.NSTEMI: lateral wall & 16 \\
			\hline
			&\\[-1em]
			Undefined ischemia/scar/supp.NSTEMI: septal & 5 \\
			\hline
			&\\[-1em]
			Undefined ischemia/scar/supp.NSTEMI: inferior wall & 3 \\
			\hline
			&\\[-1em]
			Undefined ischemia/scar/supp.NSTEMI: posterior wall & 4 \\
			\hline
			&\\[-1em]
			Undefined ischemia/scar/supp.NSTEMI: apical & 11 \\
			\hline
			&\\[-1em]
			\multicolumn{1}{c|}{Non-specific repolarisation abnormalities} &{Number of subjects} \\
			\hline
			&\\[-1em]
			Anterior wall & 18 \\
			\hline
			&\\[-1em]
			Lateral wall & 13 \\
			\hline
			&\\[-1em]
			Septal & 15 \\
			\hline
			&\\[-1em]
			Inferior wall & 19 \\
			\hline
			&\\[-1em]
			Posterior wall & 9 \\
			\hline
			&\\[-1em]
			Apical & 11 \\
			\hline
			&\\[-1em]
			\multicolumn{1}{c|}{Other states} &{Number of subjects} \\
			\hline
			&\\[-1em]
			Early repolarization syndrome & 9 \\
			\hline
		\end{tabular}
        \end{adjustbox}
		\label{table5}
	%\end{adjustwidth}
\end{table}

\begin{figure*}
	\centering
	\includegraphics[width=\linewidth]{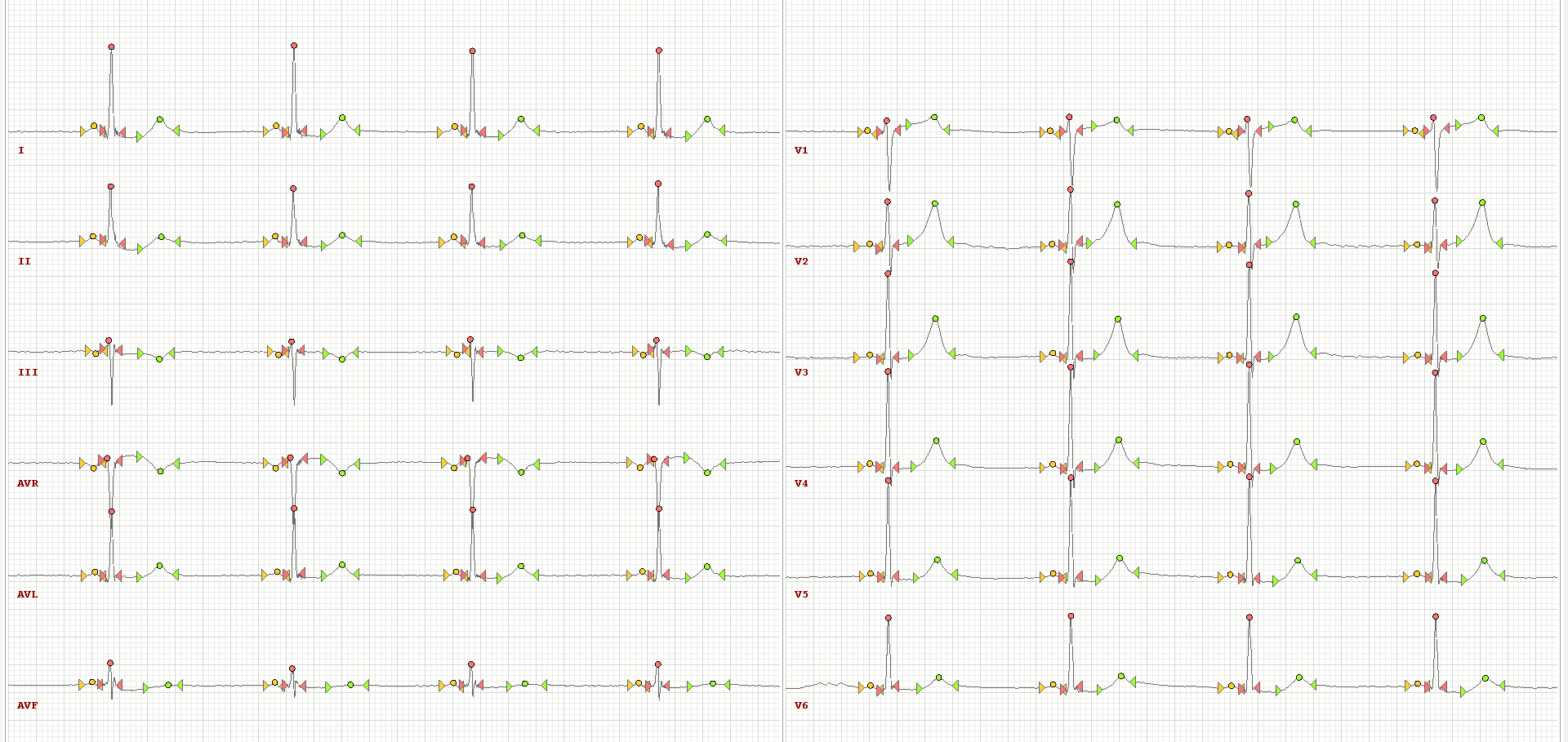}
	\caption{\NewCorrection{Example of ECG from LUDB, id=1, age: 51, sex: F. Yellow color corresponds to P waves, red to QRS complexes, green to T waves. The symbol $\rhd$ means the onset of a wave, $\circ$ means the wave peak, $\lhd$ corresponds to the offset of a wave.
Sinus rhythm.  Sinus bradycardia. Electric axis of the heart: left axis deviation. Left ventricular hypertrophy. Left ventricular overload. Non-specific repolarization abnormalities: posterior wall.}}
	\label{fig_50436612}
\end{figure*}

\begin{figure*}
	\centering
	\includegraphics[width=\linewidth]{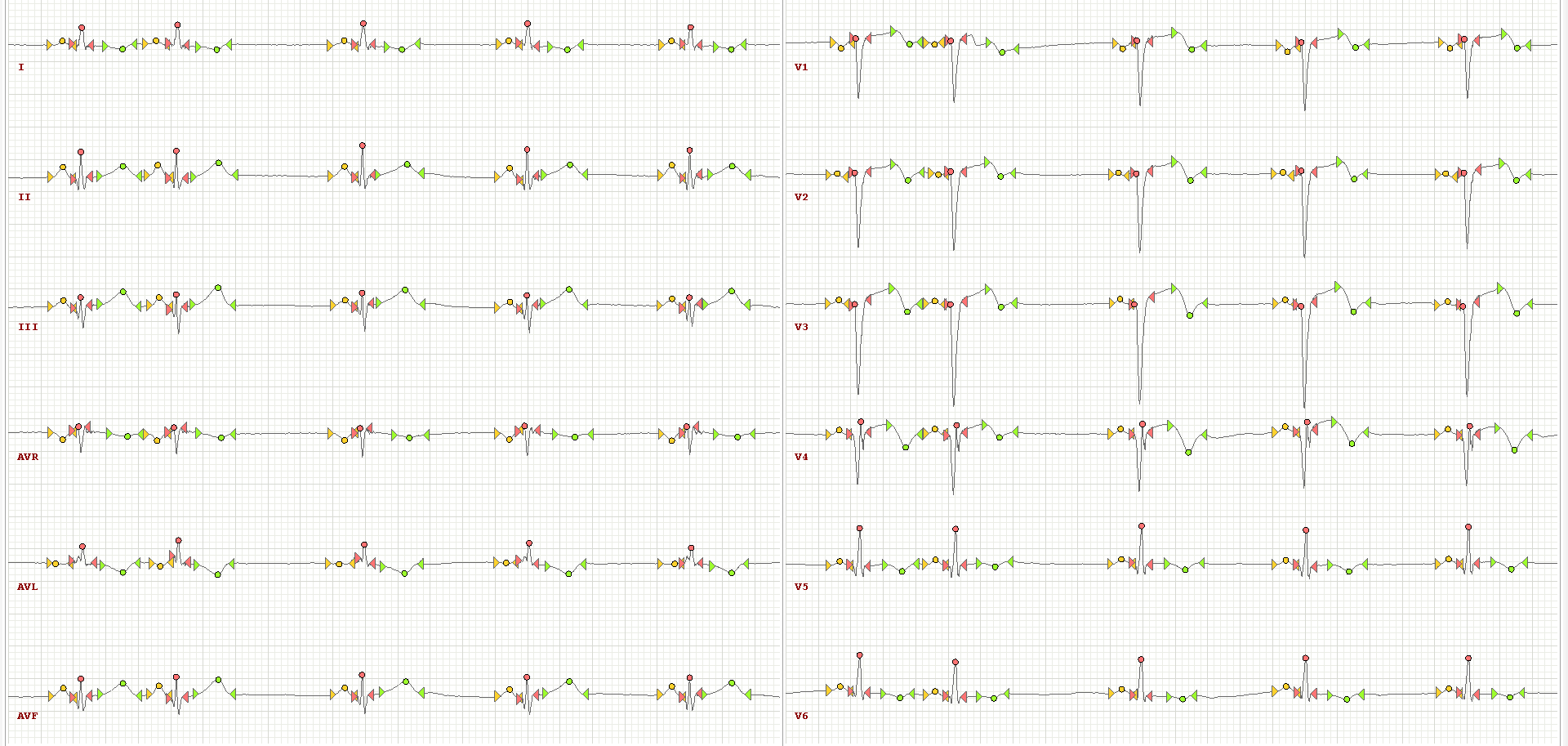}
	\caption{\NewCorrection{id = 7, age: 50, sex: M. Sinus rhythm. Electric axis of the heart: horizontal. Atrial extrasystole: SA-nodal extrasystole. Atrial extrasystole, type: single PAC. Left atrial hypertrophy. Right atrial overload. Left ventricular hypertrophy. STEMI: anterior wall. STEMI: lateral wall. STEMI: septal. STEMI: apical.
	}}
	\label{fig_50437056}
\end{figure*}

\begin{figure*}
	\centering
	\includegraphics[width=\linewidth]{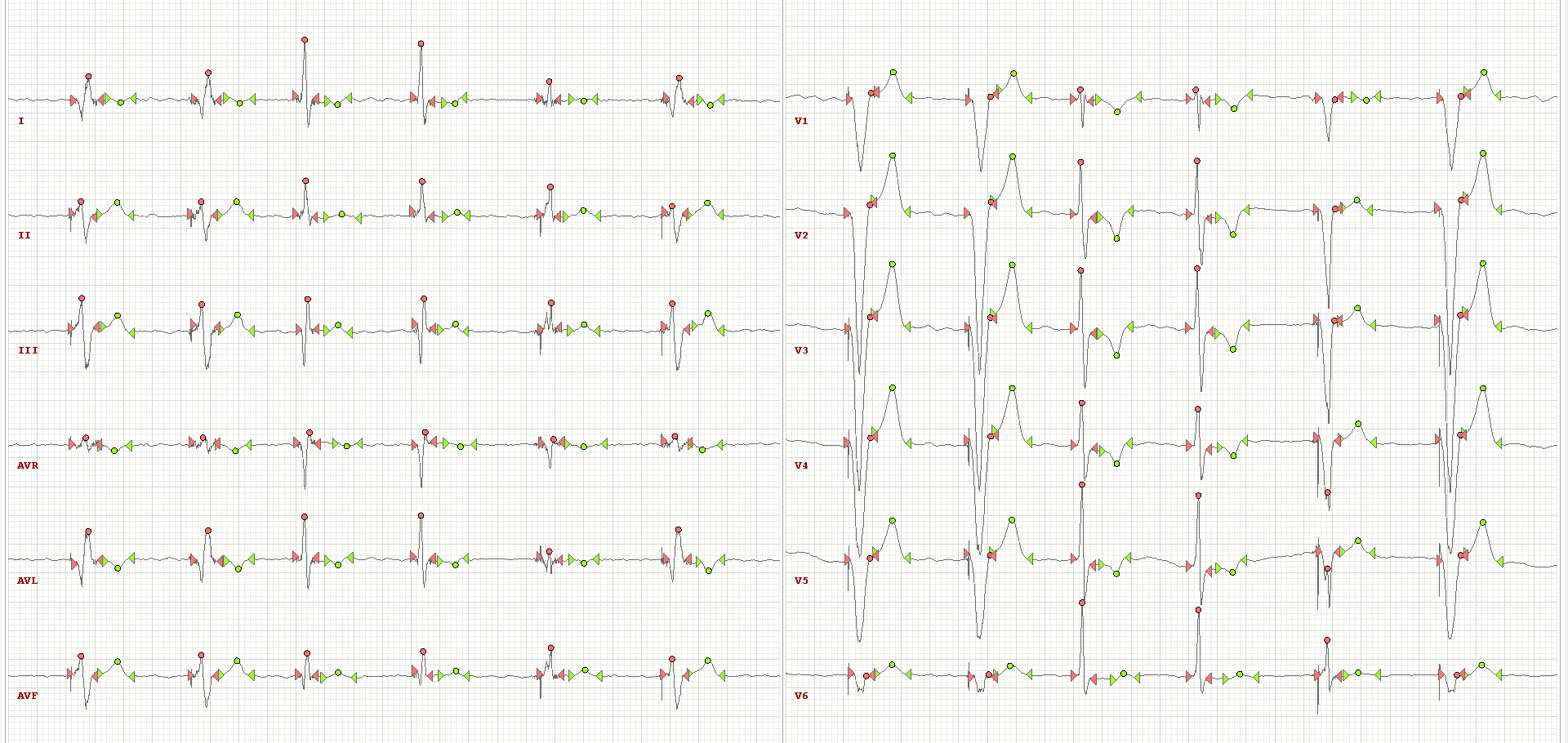}
	\caption{\NewCorrection{id = 8, age: 57, sex: F. Atrial fibrillation. Electric axis of the heart: left axis deviation. Undefined ischemia/scar/supp.NSTEMI: anterior wall. Undefined ischemia/scar/supp.NSTEMI: lateral wall. Undefined ischemia/scar/supp.NSTEMI: septal. Undefined ischemia/scar/supp.NSTEMI: apical. Pacemaker presence, undefined. UNIpolar ventricular pacing.
	}}
	\label{fig_50437173}
\end{figure*}

\begin{figure*}
	\centering
	\includegraphics[width=\linewidth]{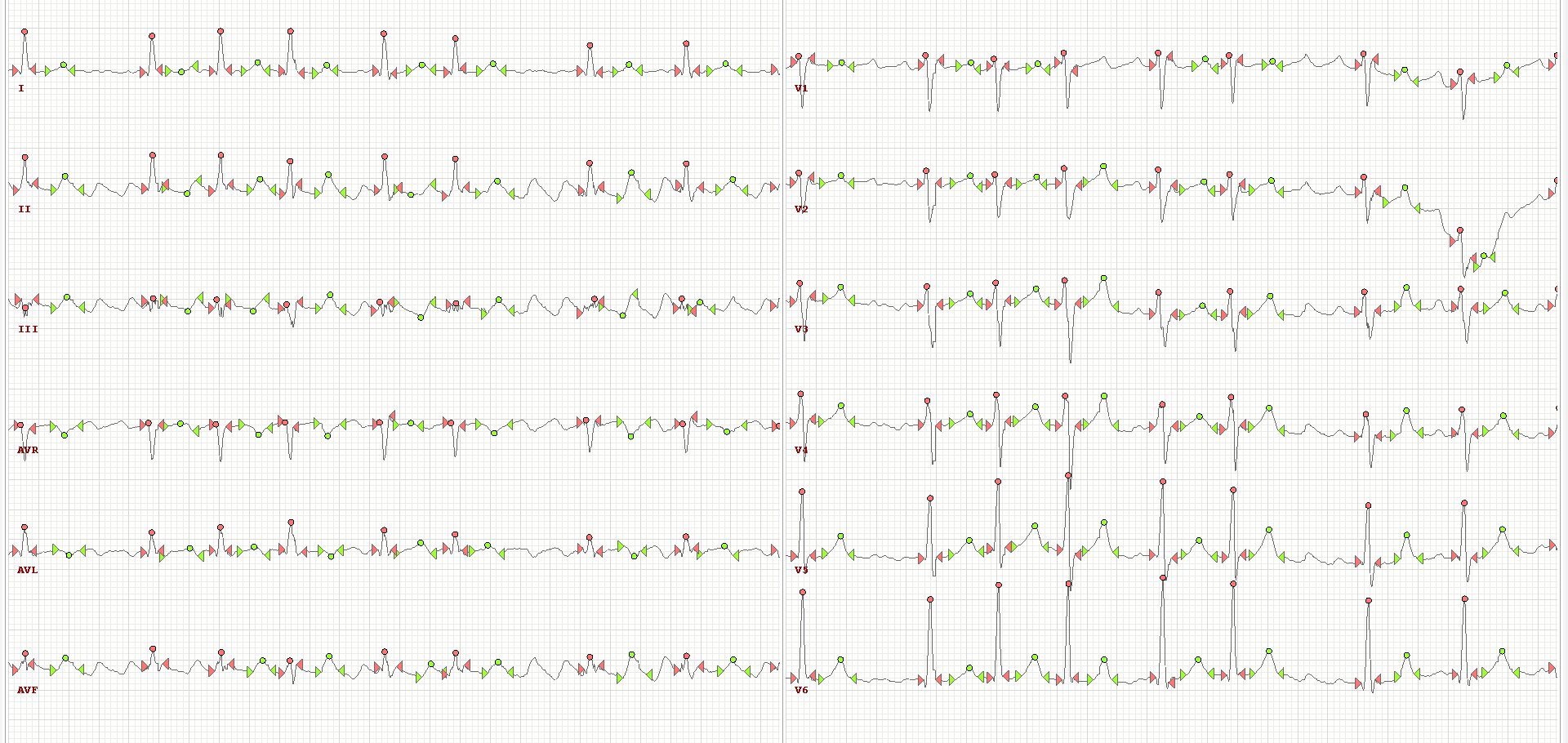}
	\caption{\NewCorrection{id = 103, age: 69, sex: M. Atrial flutter, typical. Electric axis of the heart: horizontal. Non-specific intravintricular conduction delay. Left ventricular hypertrophy.
	}}
	\label{fig_50656252}
\end{figure*}

\begin{figure*}
	\centering
	\includegraphics[width=\linewidth]{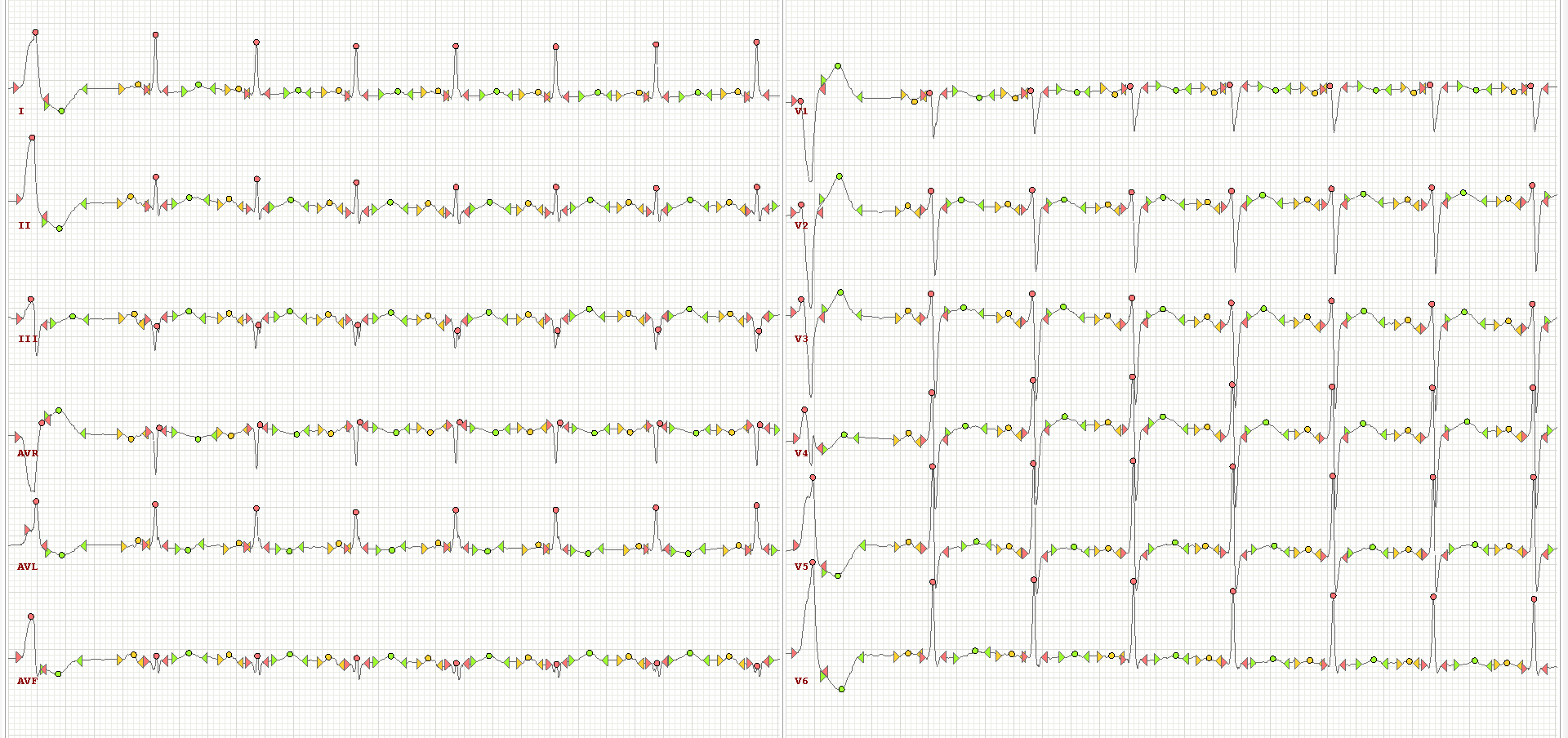}
	\caption{\NewCorrection{id = 106, age: 64, sex: M. Sinus rhythm. Electric axis of the heart: left axis deviation. Ventricular extrasystole, localisation: RVOT, antero-septal part. Ventricular extrasystole, type: single PVC. Left atrial hypertrophy. Left ventricular hypertrophy. Scar formation: posterior wall. Non-specific repolarization abnormalities: lateral wall.
	}}
	\label{fig_50665484}
\end{figure*}

\NewCorrection{Examples of ECG with manual annotations are on Figures \ref{fig_50436612}--\ref{fig_50665484}.}

%\begin{figure*}
%	\centering
%	\includegraphics[width=\linewidth]{50436790.PNG}
%	\caption{\NewCorrection{id = 4, age: 56, sex: M. Electric axis of the heart: left axis deviation. Incomplete right bundle branch block. Left atrial hypertrophy. Left ventricular hypertrophy. Ischemia: inferior wall. Scar formation: inferior wall. Undefined ischemia/scar/supp.NSTEMI: inferior wall.
%	}}
%	\label{fig_50436790}
%\end{figure*}

%\begin{figure*}
%	\centering
%	\includegraphics[width=\linewidth]{50531390.PNG}
%	\caption{\NewCorrection{id = 56, age: 46, sex: F. Sinus rhythm. Electric axis of the heart: vertical.
%	}}
%	\label{fig_50531390}
%\end{figure*}

%\begin{figure*}
%	\centering
%	\includegraphics[width=\linewidth]{50612805.PNG}
%	\caption{\NewCorrection{id = 70, age: 52, sex: M. Sinus tachycardia. Incomplete right bundle branch block. Left atrial overload.
%	}}
%	\label{fig_50612805}
%\end{figure*}

\section{Delineation algorithms}\label{algorithm}

\begin{figure}
    \centering
	\includegraphics*[width=1\linewidth]{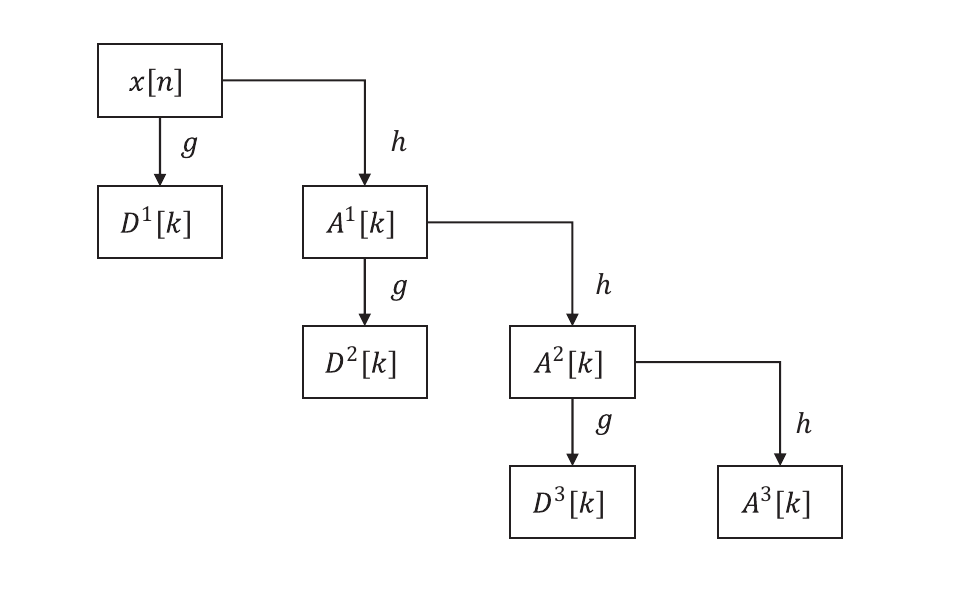}
	
	\caption{Filter bank for a discrete wavelet transform. General scheme for DWT decomposition.}
	\label{fig2}
\end{figure}

Testbed delineation tools \cite{ECGKit,Kalyakulina2018} belong to the family of methods based on discrete wavelet transform (DWT) \cite{Martinez2004, Addison2005, DiMarco2011, Bote2017}, that stems from the pioneering work by Li \cite{Li1995}. 
Commonly, a single-lead ECG signal $x[n]$ is decomposed into different frequency components by means of standard filters, Daubechies, Coiflet or biorthogonal wavelets, to name a few, as follows:

\begin{equation}
A\left[k\right] = \sum_{n} x\left[n\right] \cdot h\left[2k-n\right],
\end{equation}	

\begin{equation}
	D\left[k\right] = \sum_{n} x\left[n\right] \cdot g\left[2k-n\right],
\end{equation}	
where $h\left[n\right]$ is the low-pass filter, $g\left[n\right]$ is the high-pass filter, $D\left[k\right]$ and $A\left[k\right]$ are the resulting approximation coefficients, respectively. A more detailed representation of the frequency content of ECG signals is obtained by repeated DWT, applied to approximation coefficients, calculated at the previous round, according the general scheme shown in the Fig. \ref{fig2}.

The popular ecg-kit tool \cite{ECGKit} is based on a single-lead delineation scheme \cite{Martinez2004}. In the following we discuss the solutions of \cite{Kalyakulina2018} that allow for improving delineation accuracy of all waves and complexes, in particular, P and T waves. A \NewCorrection{comprehensive} analysis of multi-lead recordings and error correction procedures stand central here.  

\begin{figure}
    \centering
	\includegraphics*[width=1\linewidth]{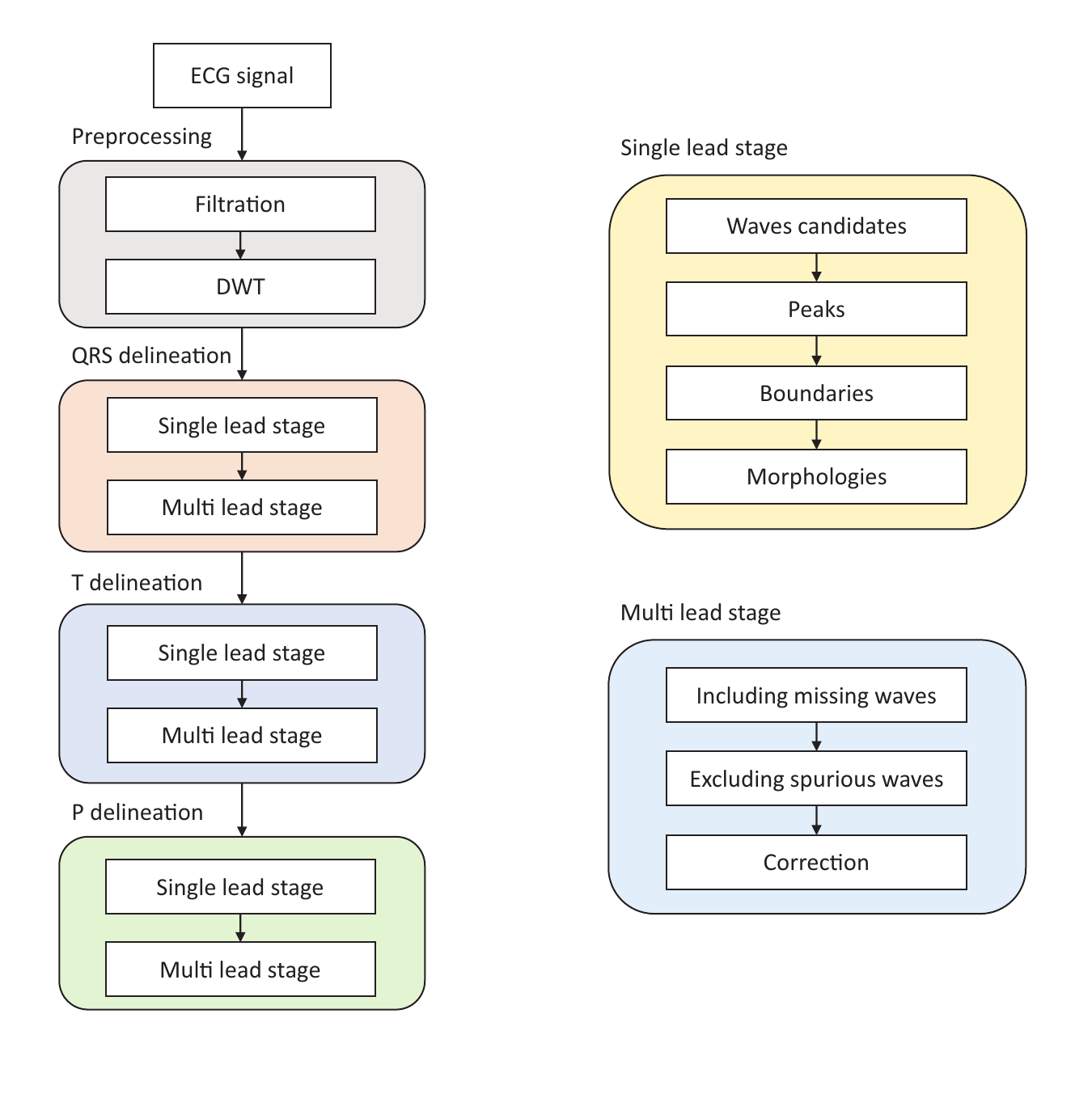}
	
	\caption{{General scheme of the ECG delineation algorithm.}
		(left) Main pipeline of the delineation algorithm consists of four stages, starting from the raw ECG signal. (right) Description of delineation steps, used for QRS, T and P waves.}
	\label{fig3}
\end{figure}

The developed delineation method consists of several stages. Delineation of each type of waves is first implemented for all ECG leads independently, and in particular order. Then, the results are refined by aggregating and comparative processing of signals from all leads. The general scheme of the algorithm is outlined in the Fig.\ref{fig3}.

The algorithm receives a raw ECG signal as an input, that is first preprocessed. Bandpass filtering removes the baseline drift and the high-frequency noise that can be caused by the muscle tone, interference from electrical appliances, poor contact between electrodes and skin, etc. Next, a discrete wavelet transform is applied to the filtered signal, yielding a set of detailed coefficients at different frequency scales. The following analysis relies on these sets obtained for ECG from each lead. 

Identifying waves and complexes of the ECG signal takes place in a specific order: QRS complex, T-wave, and then P-wave. QRS complex is detected first, since it typically has the largest amplitude, which simplifies the task. Then, T-wave is located, as its amplitude is usually greater than that of P-wave. Delineation of P-wave is viewed as the most complex task by both the cardiologists and mathematicians \cite{DiMarco2011, Martinez2004}. The amplitude of this wave often compares to noise or flutter, so that a quality detection procedure has to rely on restricting the temporal interval of interest from both sides, by QRS complex and T-wave.

\begin{figure}
    \centering
	\includegraphics*[width=1\linewidth]{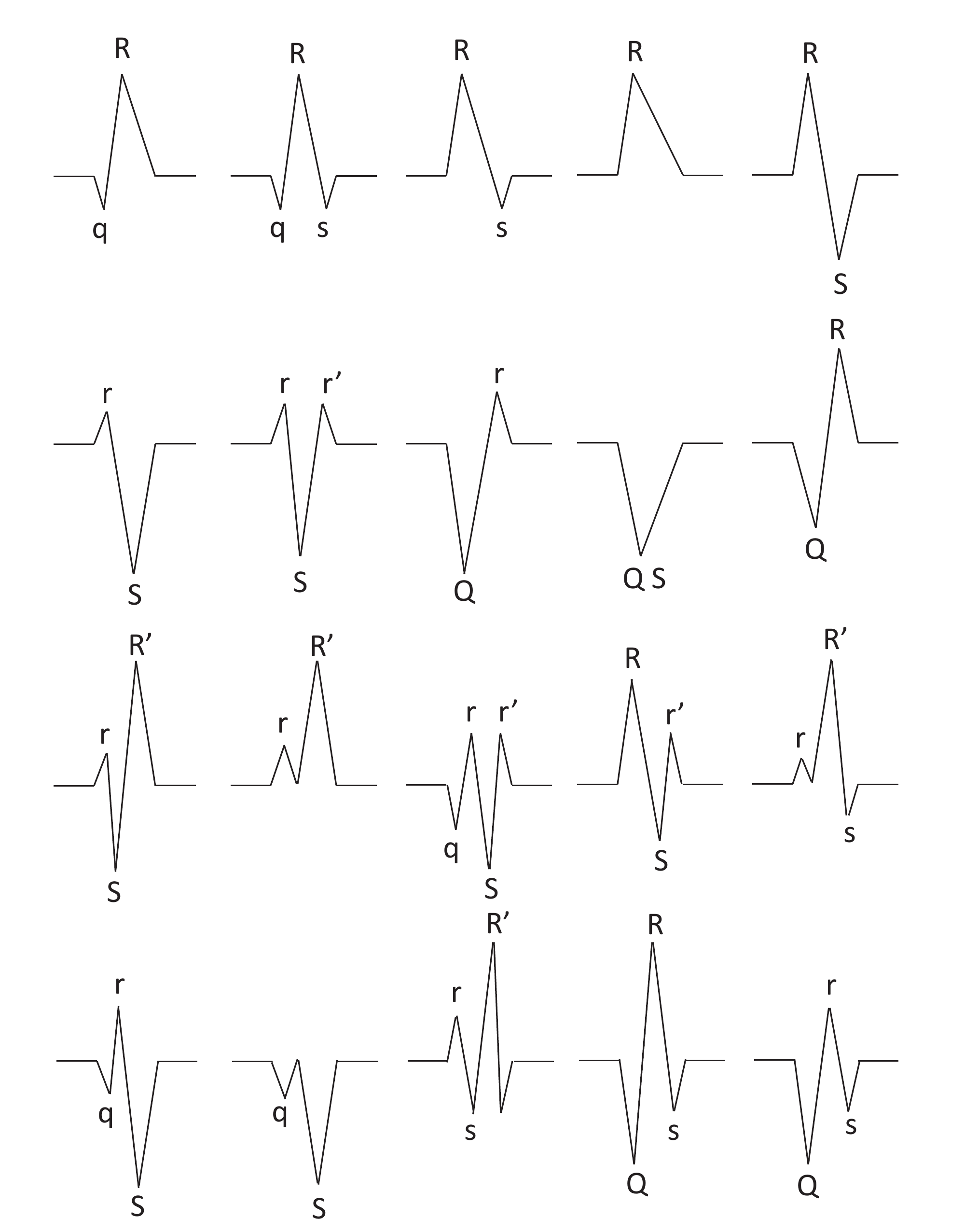}
	\caption{{Examples of QRS complex morphologies present in LUDB.}
		There are many different morphologies of the QRS complex, which can indicate the presence of various cardiovascular diseases. Their classification constitutes a challenge for automatic delineation.}
	\label{fig4}
\end{figure}

Processing each type of wave has a similar pipeline. First, the algorithm explores ECG signal from each lead separately. It selects the best candidates for the corresponding wave, then determines its peak and boundaries. The algorithm by Kalyakulina et al. \cite{Kalyakulina2018} implements yet another feature, classifying the morphology of the detected wave by determining reference points (onsets, peaks, ends). Matching them to model cases gives a much more advanced diagnostic information than duration and amplitude values would offer. The particular morphologies of the QRS complex, recognized by the algorithm, are shown in the Fig. \ref{fig4}. Orientation of the complex, its extremal points, the number of additional peaks or, conversely, the lack of some, are key to the diagnostic process, detecting cardiac arrhythmias or the presence of cardiovascular diseases.

After all waves of a certain type are found for the outputs from all leads, the algorithm performs a comparative analysis, aimed at correcting omissions or spurious waves, appearing in recordings for certain leads. As a formal validity threshold for a complex occurrence, we require its presence in at least 8 out of 12 leads. That is, if for some heartbeat the T-wave is detected for 10 leads out of 12, then it is taken that this wave is also present for the other two leads. Conversely, if the complex is found in \NewCorrection{at most} one third of the total number of leads, then it is retracted from delineation. \NewCorrection{We don’t use the multilead correction if the complex was detected on 5\dots8 leads.} Additionally, averaging the times of the corresponding reference points for the matching complexes across the leads reduces the effect of noise and other disturbances. After this multi-lead correction, delineation steps down to the subsequent wave, taking an advantage of adjusted locations of preceding waves.

\begin{figure*}
    \centering
	\includegraphics*[width=160mm]{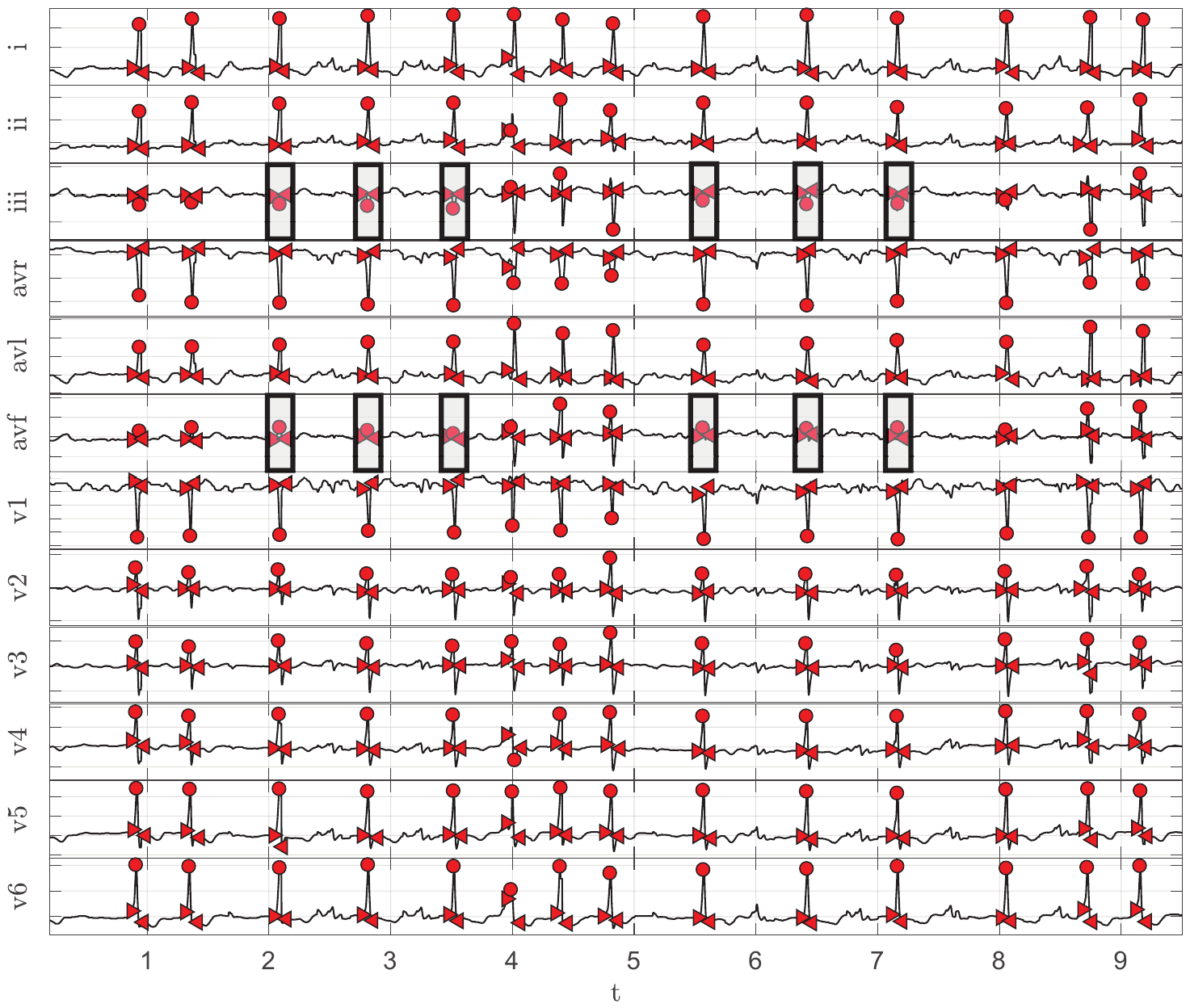}
	
	\caption{{Multi-lead refinement of delineation.}
		Gray frames show the complexes, which fall short of the single lead analysis, but are recovered by the multi-lead refinement. \NewCorrection{For each missed complex the averaged value for the start and the averaged value for the end of the complex were found. The averaging is performed over those leads, where the complex is found. The global extremum in this interval (from the average start to the average end) is considered as the peak of the complex.}}
	\label{fig5}
\end{figure*}

Instructively, some failures in the single-lead signal processing are apparently due to alternating morphologies of a complex in the ECG signal, which the adaptive detection threshold does not follow efficiently enough \cite{Kalyakulina2018}. However, when the complexes are missed in less than one third of leads, their delineation is also restored by the multi-lead analysis, as exemplified in Fig. \ref{fig5}, and a corresponding morphological anomaly is noted down.

\section{Algorithm validation}\label{results}

We validate the described tools \cite{ECGKit,Kalyakulina2018} with two open access datasets, the newly introduced LUDB and QTDB \cite{qtdb}, both manually annotated by cardiologists, but distinct in the number of leads (12 and 2, respectively), number of subjects (200 and 105) and duration of recordings (10 and 15 seconds). The reference points of complexes found by an automated delineation are checked against the manually marked ones, the chosen tolerance window interval of 150 ms complies with ANSI/AAMI-EC57:1998 standard \cite{standart1999}.

When an algorithm determines a point correctly (i.e. within the 150 ms interval of a manual point), it is counted as true positive (TP). Likewise, when a point suggested by the algorithm is absent in the manual markup, the case is counted as false positive (FP). If the algorithm fails to identify the point, which is present in the database, the case is false negative (FN). For TP cases one also calculates a time mismatch between the automated and manually assigned locations, and this quantity is referred to as ``error''. The quality of the algorithm is characterized by the following four metrics, implemented in \cite{Rincon2011, DiMarco2011, Martinez2004, Bote2017}: average error $m$, its standard deviation $\sigma$, sensitivity $Se(\%) = TP / (TP + FN)$, and positive predictive value (precision) $PPV(\%) = TP / (TP + FP)$. For Kalyakulina et al. method, all these quantities are computed based on the set that is pooled from the point-to-point match analysis in each single lead. 

Table \ref{table6} summarizes the assessment of the two tools \cite{ECGKit,Kalyakulina2018} against LUDB an QTDB, and gives validation data for the other methods against QTDB, borrowed from the literature 
\cite{Martinez2004, Rincon2011, DiMarco2011, Bote2017}, and against LUDB \cite{Chen2020}.

\begin{table*}
	\caption{Quality of delineation algorithms validated on LUDB and QTDB.
             Best values of $Se$ and $PPV$ for each key point on QTDB are in bold.}
	\label{table6}
	\medskip
\begin{adjustbox}{width=1\linewidth}
	$
\begin{array}{|c|c|c|c|c|c|c|c|c|}
\hline		
\multicolumn{2}{|c|}{ } & \text{P onset} & \text{P peak} & \text{P offset} & \text{QRS onset} & \text{QRS offset} & \text{T peak} & \text{T offset} \\
\hline		
               & Se(\%)    &   98.46     & 98.46     &  98.46      &   99.61    &  99.61     & 99.03      &98.03       \\
\raisebox{1.4ex}[0cm][0cm]{\text{Kalyakulina {\it et al.} \cite{Kalyakulina2018}}} 
               & PPV(\%)   &   96.41     & 96.41     &  96.41      &   99.87    &  99.87     & 98.84      &98.84       \\
\raisebox{1.35ex}[0cm][0cm]{(LUDB)}
           & m\pm\sigma(ms) &-2.7\pm10.02 &-0.3\pm6.2 & -0.4\pm11.4 & -8.1\pm7.7 & 3.8\pm8.8  &4.0\pm7.4   &5.7\pm15.5  \\
\hline                                                                         
                & Se(\%)    &   97.46     & 97.50     &  97.53      &   98.42    &  98.42     & 98.24      &96.16       \\
\raisebox{1.4ex}[0cm][0cm]{\text{Kalyakulina {\it et al.} \cite{Kalyakulina2018}}} 
                & PPV(\%)   &\text{\bf 97.86}  &\text{\bf 97.89} &\text{\bf 97.93} &   98.24    &  98.24     & 98.24      &94.87       \\
\raisebox{1.35ex}[0cm][0cm]{(QTDB)}
                & m\pm\sigma(ms) &-3.5\pm13.8   &4.3\pm10.0 & 3.4\pm12.7  &-5.1\pm6.6  &4.7\pm9.5   &7.2\pm13.0  &13.4\pm18.5 \\
 \hline
               & Se(\%)    &   98.43     & 98.43 &    98.43    &   99.89    &   99.89    & 99.27      &  99.21     \\
\text{Chen {\it et al.} \cite{Chen2020} (LUDB)}
               & PPV(\%)   &    96.44    & 96.44 &    96.44    &   99.86    &   99.86    & 98.85      &  98.85     \\
            &m\pm\sigma(ms)& 2.2\pm7.4  &-0.76\pm5.5 &-6.5\pm10.7&15.4\pm14.6&-3.8\pm13.6&-0.5\pm5.5 &-1.2\pm6.8\\
 \hline
& Se(\%)    &   88.26     & 89.64 &    91.08    &   99.52    &   99.51    & 85.62      &  85.00     \\
\text{ecg-kit \cite{ECGKit} (LUDB)}
& PPV(\%)   &    82.43    & 83.73 &    85.07    &   91.36    &   91.35    & 94.91      &  94.22     \\
&m\pm\sigma(ms)& 16.2\pm31.7&12.0\pm31.1 & 7.9\pm22.3&-3.33\pm14.3&3.7\pm15.9&11.9\pm32.1 &-3.4\pm32.8\\
\hline                                                                         
               & Se(\%)    &   98.64     & 98.64 &    98.64    &   99.60    &   99.60    & 96.86      &  96.86     \\
\text{ecg-kit \cite{ECGKit} (QTDB)}
               & PPV(\%)   &    70.75    & 70.75 &    70.75    &   91.33    &   91.33    & 91.52      &  91.52     \\
             &m\pm\sigma(ms)&-0.5\pm12.0&9.4\pm1.1 & -2.7\pm7.9&-5.7\pm3.9  &-0.5\pm9.8  &2.2\pm6.3   &-0.6\pm8.2  \\
\hline                                                                          
               & Se(\%)    &   98.12     & 99.15     &  99.87      &   99.50    &  99.50     & 99.41      &96.98        \\
\raisebox{1.4ex}[0cm][0cm]{\text{Bote \textit{et al.} \cite{Bote2017}}} 
               & PPV(\%)   &   94.26     & 95.11     &  96.03      &\text{\bf 99.78} &\text{\bf 99.78} &\text{\bf 98.96} &95.98        \\
\raisebox{1.35ex}[0cm][0cm]{(QTDB)}
& m\pm\sigma(ms) &23.9\pm19.5   &13.8\pm8.8 &-1.9\pm10.4  & 6.4\pm5.5  &-5.2\pm10.8 &9.0\pm15.4  &-12.9\pm18.6 \\
\hline                                                                          
               & Se(\%)    &   98.15     & 98.15     &  98.15      &\text{\bf 100.00} & \text{\bf 100.00} & 99.72      & 99.77       \\
\raisebox{1.4ex}[0cm][0cm]{\text{DiMarco \textit{et al.} \cite{DiMarco2011}}} 
               & PPV(\%)   &   91.00     & 91.00     &  91.00      & \text{--}  &  \text{--} & 97.76      & 97.76       \\
\raisebox{1.35ex}[0cm][0cm]{(QTDB)}
& m\pm\sigma(ms) & -4.5\pm13.4 &-4.7\pm9.7 &-2.5\pm13.0  &5.1\pm7.2   &0.9\pm8.7   &-0.3\pm12.8 & 1.3\pm18.6  \\
\hline                                                                          
               & Se(\%)    &   98.87     &\text{\bf 99.87}   &  98.75      &  99.97     & 99.97      &\text{\bf 99.97}  & 99.77       \\
\raisebox{1.4ex}[0cm][0cm]{\text{Martinez \textit{et al.} \cite{Martinez2004}}} 
               & PPV(\%)   &   91.03     & 91.03     &  91.03      & \text{--}  &  \text{--} & 97.79      &97.79        \\
\raisebox{1.35ex}[0cm][0cm]{(QTDB)}
& m\pm\sigma(ms) &2.0\pm14.8   &3.6\pm13.2 & 1.9\pm12.8  & 4.6\pm7.7  & 0.8\pm8.7  &0.2\pm13.9  & -1.6\pm18.1 \\
\hline                                                                          
               & Se(\%)    &\text{\bf 99.87}  &\text{\bf 99.87}    &\text{\bf 99.91} & 99.97      & 99.97      &\text{\bf 99.97} &\text{\bf 99.97}  \\
\raisebox{1.4ex}[0cm][0cm]{\text{Rincon \textit{et al.} \cite{Rincon2011}}} 
               & PPV(\%)   &   91.98     & 92.46     &  91.70      & 98.61      & 98.72      &98.91       &\text{\bf 98.50}     \\
\raisebox{1.35ex}[0cm][0cm]{(QTDB)}
        & m\pm\sigma(ms)   &8.6\pm11.2   &10.1\pm8.9 & 0.9\pm10.1  & 3.4\pm7.0  &3.5\pm 8.3  &3.7\pm13.0  & -2.4\pm16.9 \\
\hline                                                                          
\multicolumn{2}{|c|}{2\sigma_{\text{CSE}}\,(ms)}
                           &    10.2     & \text{--} &  12.7       & 6.5        & \text{\NewCorrection{11.6}}        &  \text{--} & 30.6       \\
\hline
\end{array}      
$
\end{adjustbox}
\end{table*}

In result, for both LUDB and QTDB, the sensitivity values for the onsets and peaks of the P, QRS and T waves are above 97\%, and the standard deviation $\sigma$ is \NewCorrection{almost} within the limits set by the standard \cite{standart1985}: \NewCorrection{it must be at most $2\sigma_{\text{CSE}}$}. \NewCorrection{The exceptions are the P wave onset for QTDB, where $\sigma$ is 3 ms larger, and QRS onset for both databases, where  $\sigma$ is 1.2 ms larger for LUDB and 0.1 ms larger for QTDB}. The maximal error is observed for the T-wave offset, whose delineation is a well-known hard problem, both from the mathematical and for the cardiological perspectives \cite{Mehta2008}. For the QRS complex, a relatively simple task, the performance of all methods is next to perfect, with occasionally slightly worse rate for the method by Kalyakulina at al. The more challenging task of detecting P and T waves is performed also almost equally well by all methods on QTDB, but the method by Kalyakulina et al. substantially outperforms ecg-kit for LUDB. This is an anticipated result, since the former method takes the full advantage of LUDB 12-lead format, that allows to reduce detection failures and appearance of spurious complexes, and to improve an accuracy of timing the key points by the multi-lead refinement of delineation. 

\NewCorrection{QTDB can be used to validate different methods for ECG delineation, as well as to train new deep learning algorithms for delineation. We believe that architectures like U-net \cite{Ronneberger2015} will allow achieving better results than known algorithms. For some preliminary results from using QTDB to train U-net-like network, see \cite{Moskalenko2019}.}

\section{Conclusion}\label{conclusions}

Despite an urgent need in thoroughly annotated and open datasets of human ECGs to serve testbeds for delineation algorithms, the offer remains quite limited \cite{mitdb,edb,qtdb}. Moreover, each case comes short of having multi-lead recordings, a standard output for modern hospital cardiographs,  and a manual expert markup of all kinds of waves (P, QRS, and T). Ideally, the recordings would be supplied with diagnosis or a note on abnormalities in ECG, that additionally enables training and validating the algorithms for an automated identification of possible pathology. 

The presented Lobachevsky University Database is a step to fill the existing gap. Openly accessible at Lobachevsky University website and \NewCorrection{available on} PhysioNet \cite{Our_PhysioNet}, it contains 12-lead ECG recordings for 200 subjects (hospital patients and participants without a history of complaints) in wfdb (PhysioNet) format, manually annotated \NewCorrection{(except for U-waves)} and \NewCorrection{supplied} with noticed abnormalities. Moreover, it offers a variety of complex morphologies to challenge delineation algorithms. A case study that employed ecg-kit \cite{ECGKit} and our recently developed delineation algorithm \cite{Kalyakulina2018} demonstrates how one can take a full advantage of multi-lead recordings to implement error corrections in signals from separate leads, and improve recognition of complex wave morphologies, as well as precision of timing for delineation points, as compared to the performance on the 2-lead dataset. The further \NewCorrection{extension} of LUDB, that would not simply enrich the base, but will make it suitable for exploring machine learning and neural network algorithms for an automated diagnosis, is to follow. It would be also important to receive independent manual delineations by the other experts.

 Our results confirm that some delineation tools can have a considerably different performance on different datasets. Different instrumental origin of ECG is only one, and probably a minor reason for that. The inevitable variability in individual expert opinion on delineation and diagnosis could give a much greater impact, both at the validation stage and for the end use. However, one still lacks enough data to evaluate and \NewCorrection{accommodate} this issue. \NewCorrection{Admittedly}, the future quality assurance of delineation algorithms will \NewCorrection{emphasize} the robust albeit next to perfect performance over a wealth of datasets, rather than maximizing it against a given example.

\section*{Acknowledgements}
We acknowledge the support by the Ministry of Science and Higher Education of the Russian Federation, Agreements No. 074-02-2018-330\,(1) and No. 13.1902.21.0026.

\section*{Author contributions}
M.V. Ivanchenko, N.Yu. Zolotykh, G.V. Osipov, K.A. Kosonogov and A.V. Nikolskiy conceived and supervised the study. 
V.A. Moskalenko, A.I. Kalyakulina and I.I. Yusipov performed data curation and analysis. 
M.V. Ivanchenko, N.Yu. Zolotykh, K.A. Kosonogov, A.V. Nikolskiy,  A.I. Kalyakulina and I.I. Yusipov wrote the paper.

\begin{IEEEbiography}[{\includegraphics[width=1in,height=1.25in,clip,keepaspectratio]{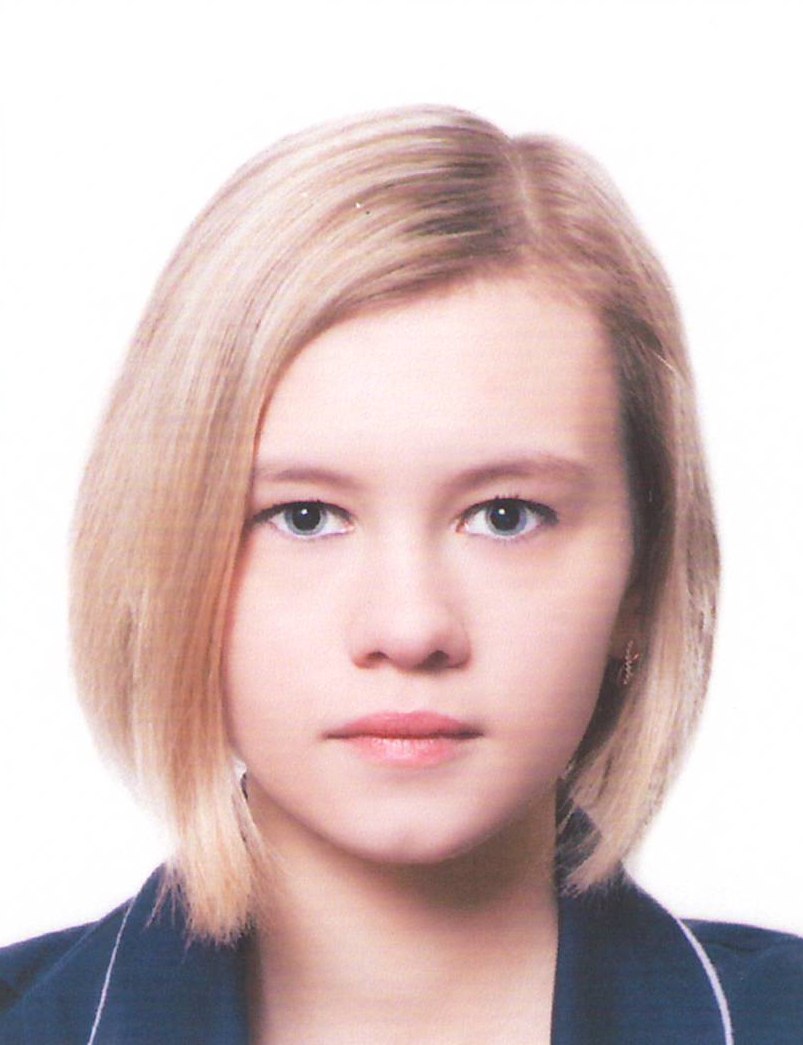}}]{A. I. Kalyakulina} graduated from Lobachevsky State University of Nizhni Novgorod in 2016. Now she is a Research Assistant at Lobachevsky State University of NIzhni Novgorod. The field of her scientific interests includes mathematical modelling of living systems, high-performance computing, nonlinear dynamics.      
\end{IEEEbiography}

\begin{IEEEbiography}[{\includegraphics[width=1in,height=1.25in,clip,keepaspectratio]{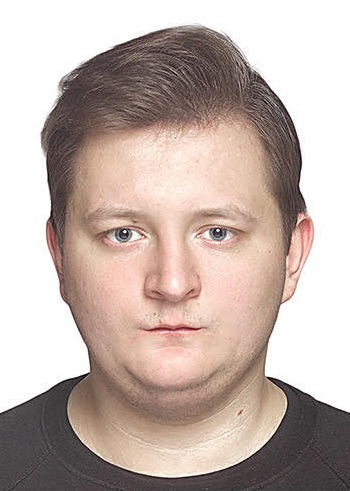}}]{I. I. Yusipov} graduated from Lobachevsky State University of Nizhni Novgorod in 2016. Now he is a research assistant at Lobachevsky State University of NIzhni Novgorod. The field of his scientific interests includes quantum physics, machine learning, high performance computing.  
\end{IEEEbiography}

\begin{IEEEbiography}[{\includegraphics[width=1in,height=1.25in,clip,keepaspectratio]{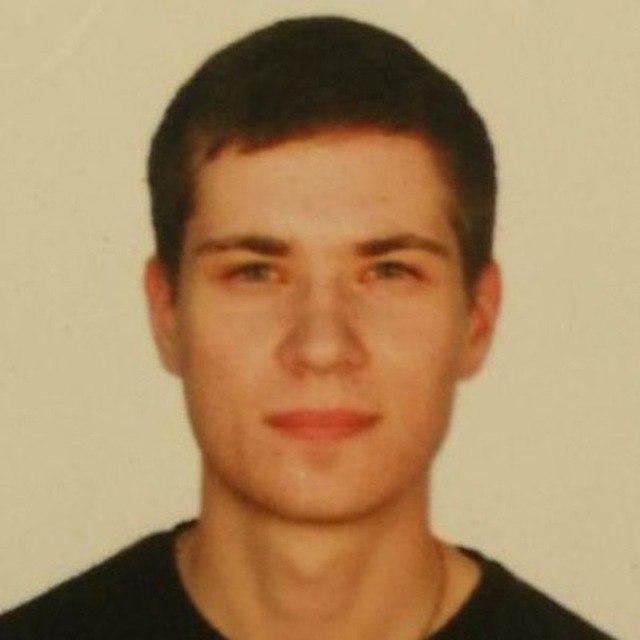}}]{V. A. Moskalenko} graduated from Lobachevsky State University of NIzhni Novgorod in 2019. Now he is a graduate student at the same University. The field of his scientific interests are machine learning.  
\end{IEEEbiography}

\begin{IEEEbiography}[{\includegraphics[width=1in,height=1.25in,clip,keepaspectratio]{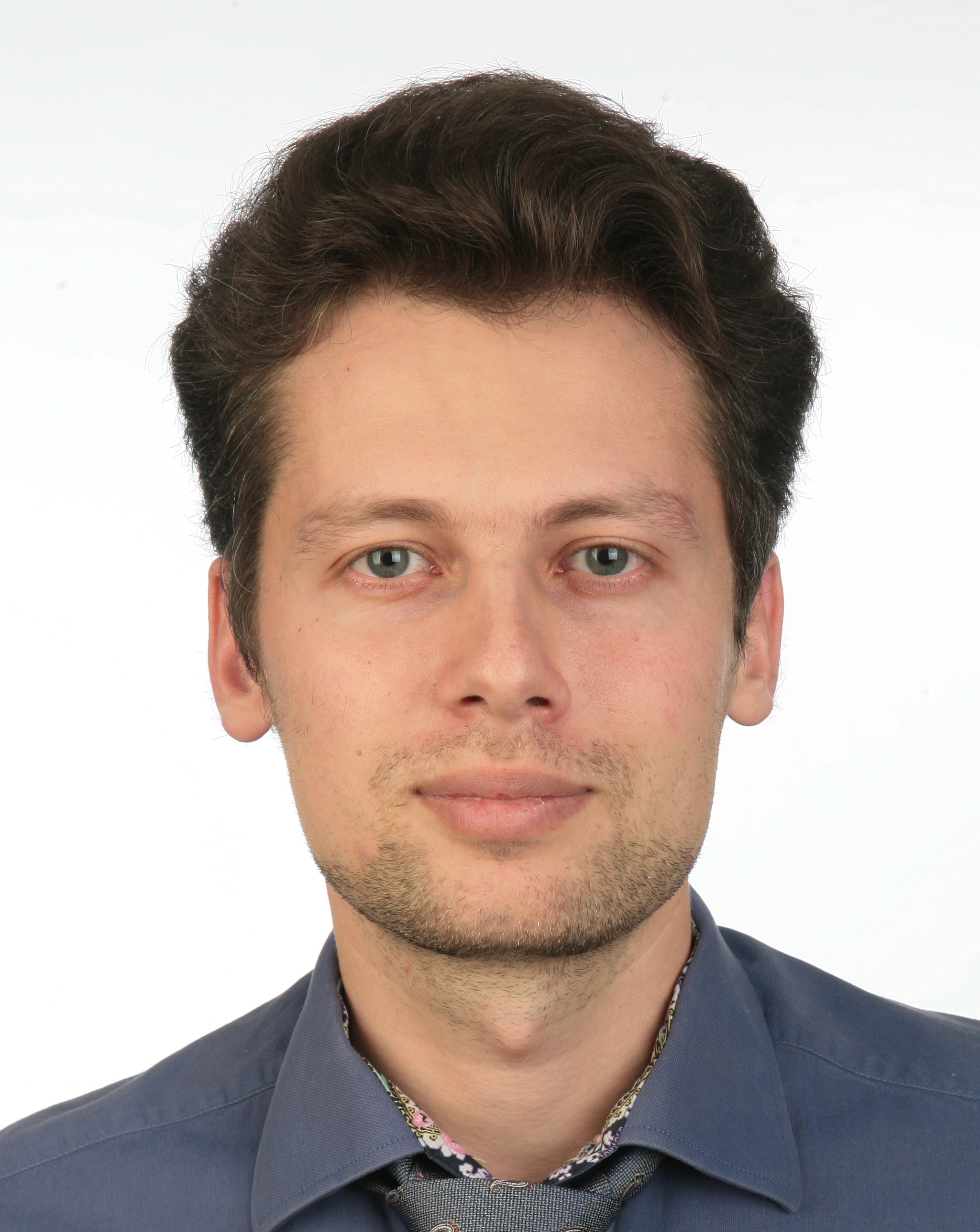}}]{A. V. Nikolskiy} graduated from faculty of medicine of State Medical Academy of Nizhni Novgorod in 2007. He received a PhD degree in Medicine in 2010. Now he is a physician of cardiovascular surgery at cardiovascular surgery department of City Clinical Hospital No 5 Nizhny Novgorod. The scope of he's practical and scientific interests includes catheter ablation of arrhythmias, surgical treatment heart rhythm disorders, arrhythmology and electrophysiology.  
\end{IEEEbiography}

\begin{IEEEbiography}[{\includegraphics[width=1in,height=1.25in,clip,keepaspectratio]{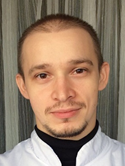}}]{K. A. Kosonogov} graduated from faculty of medicine of State Medical Academy of Nizhni Novgorod in 2007. He received a PhD degree in Medicine in 2016. Now he is a physician of cardiovascular surgery at cardiovascular surgery department of City Clinical Hospital No 5 Nizhny Novgorod. The scope of he's practical and scientific interests includes arrhythmology, cardiac electrophysiology, cardiac arrhythmia ablation, atrial fibrillation,cardiac pacing, lead extraction. 
\end{IEEEbiography}

\begin{IEEEbiography}[{\includegraphics[width=1in,height=1.25in,clip,keepaspectratio]{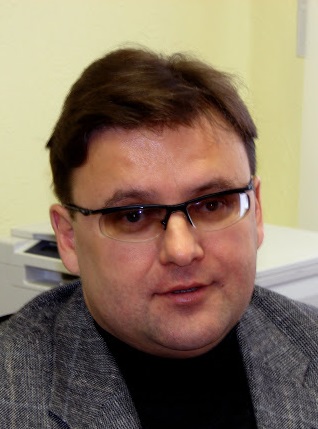}}]{G. V. Osipov} graduated from Lobachevsky State University of Nizhni Novgorod in 1982. He received a PhD degree in Physics and Mathematics in 1989, and a Doctor of Science in 2004. Now he is a professor at Lobachevsky State University of Nizhni Novgorod and the head of the Department of the Control Theory and System Dynamics. The field of his scientific interests includes nonlinear dynamics, synchronization, mathematical modeling,	controlling chaos, pattern formation, theory of bifurcations, computational neuroscience, machine learning.  
\end{IEEEbiography}

\begin{IEEEbiography}[{\includegraphics[width=1in,height=1.25in,clip,keepaspectratio]{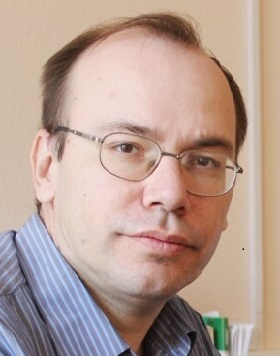}}]{N. Yu. Zolotykh} graduated from Lobachevsky State University of Nizhni Novgorod in 1995. He received a PhD degree in Mathematics in 1998, and a Doctor of Science in 2014. Now he is a professor at Lobachevsky State University of NIzhni Novgorod. The field of his scientific interests includes machine learning, computational geometry, discrete geometry, discrete optimization.  
\end{IEEEbiography}

\begin{IEEEbiography}[{\includegraphics[width=1in,height=1.25in,clip,keepaspectratio]{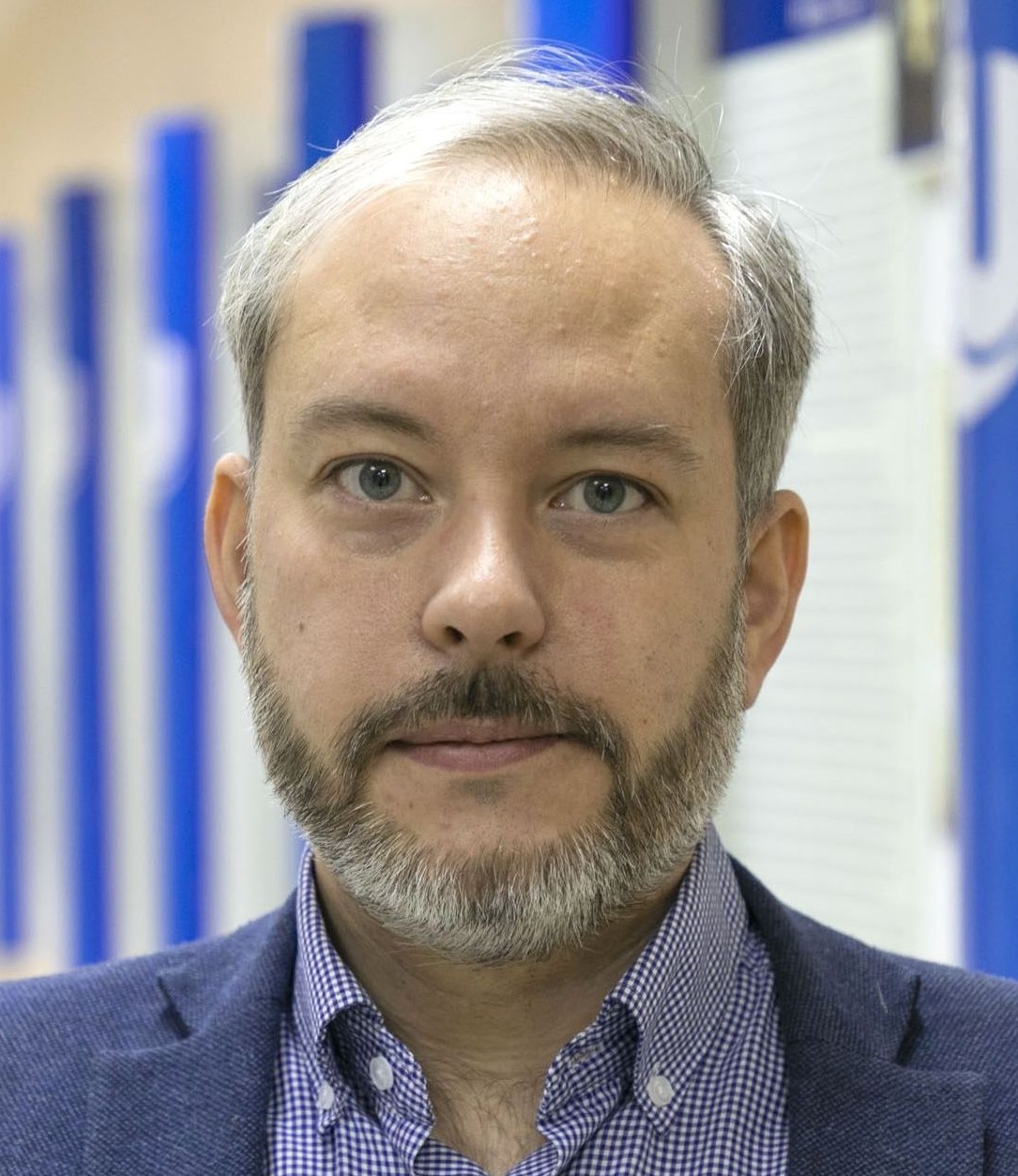}}]{M. V. Ivanchenko} graduated from Lobachevsky State University of Nizhny Novgorod in 2004. He received a PhD degree in Physics and Mathematics in 2007, and a Doctor of Science in 2012. Now he is a professor at Lobachevsky State University of Nizhny Novgorod. The field of his scientific interest includes nonlinear dynamics, complex systems, bioinformatics and data analysis.  
\end{IEEEbiography}

\EOD

\end{document}